\documentclass[usenatbib,times]{mn2e}
\usepackage{graphicx}
\usepackage{epsf}

\def\xte{{\it RXTE}}

\def\H0{{\rm ~km~s^{-1}~Mpc^{-1}}}

\def\cx1{Cyg~X-1}
\def\lone{L$_1$}
\def\ltwo{L$_2$}
\def\lthree{L$_3$}

\begin{document}

\title[Frequency-resolved spectroscopy in Cyg X-1]{Breaking the spectral degeneracies in black hole binaries with fast timing data}

\author[Axelsson \& Done]{Magnus Axelsson,$^{1,2}$\thanks{email: magnusa@astro.su.se}
and Chris Done$^{3}$\\
$^{1}$Oskar Klein Center for CosmoParticle Physics, Department of Physics, Stockholm University, SE-106 91 Stockholm, Sweden\\
$^{2}$Department of Astronomy, Stockholm University, SE-106 91 Stockholm, Sweden\\
$^{3}$Department of Physics, Durham University, South Road, Durham DH1 3LE, UK\\
}

\date{Accepted --. Received --; in original form --}

\pagerange{\pageref{firstpage}--\pageref{lastpage}} \pubyear{2018}

\maketitle

\begin{abstract}

The spectra of black hole binaries in the low/hard state are complex, with evidence for 
multiple different Comptonisation regions contributing to the hard X-rays in addition to 
a cool disc component. We show this explicitly for some of the best {\it RXTE} data from 
Cyg~X-1, where the spectrum strongly requires (at least) two different Comptonisation 
components in order to fit the continuum above 3~keV, where the disc does not contribute. 
However, it is difficult to constrain the shapes of these Comptonisation components 
uniquely using spectral data alone. Instead, we show that additional information from 
fast variability can break this degeneracy. Specifically, we use the observed variability 
power spectra in each energy channel to reconstruct the energy spectra of the variability 
on timescales of $\sim 10$\,s, $1$\,s and $0.1$\,s. The two longer timescale spectra 
have similar shapes, but the fastest component is dramatically harder, and has strong 
curvature indicating that its seed photons are not from the cool disc. We interpret this 
in the context of propagating fluctuations through a hot flow, where the outer regions 
are cooler and optically thick, so that they shield the inner region from the disc. The 
seed photons for the hot inner region are then from the cooler Comptonisation
region rather than the disc itself.

\end{abstract}

\section{Introduction}
\label{intro}
The energy spectra of the low/hard state of black hole binaries have
complex curvature, and robustly require additional components as well
as the well-known `disc plus power law'. This is shown most clearly in
spectra spanning the broadest bandpass \citep{mak08,now11}, but even in
3-100~keV data alone (e.g., from {\it RXTE}) the continuum shape has a clear
hardening beyond 10~keV which cannot be accounted for by reflection
alone \citep[unless the reflection parameters are extreme:][]{fab14}. 
\citet{now11} show that this can be modelled by several
different continuum components, which could plausibly derive from a
radial stratification of temperature and/or optical depth of the
Comptonising hot flow \citep{dis01,ibr05,mak08,yam13,bas17}. 
Alternatively, this curvature could arise from a completely
different region, potentially indicating synchrotron emission from the
jet \citep{mar05}. Conversely, the curved spectrum could arise
from a single region if the electron distribution is not completely
thermal \citep[hybrid thermal/non-thermal models:][]{pc98,
gie99, ibr05}. Plainly, spectral
fitting alone is highly degenerate, but the longer term spectral
evolution in the low/hard state, where the overall spectrum softens
with increasing luminosity, can be generally intrerpreted in models
where the inner disc evaporates into a hot flow above the innermost
stable circular orbit. Decreasing this truncation radius with
increasing mass accretion rate leads to stronger disc emission, which
leads to stronger Compton cooling and a softer spectrum, as observed
\citep[see, e.g., the review by][]{don07}. Multiple Comptonisation
regimes are quite naturally expected in this geometry as the part of
the hot flow closest to the truncated disc is more strongly
illuminated, so should have a softer spectrum than the inner parts of
the hot flow closest to the black hole. 

Here we use the additional information from spectral evolution during
fast variability in the low/hard state to determine the continuum
component shapes, and hence better constrain their physical origin.
The fast variability (timescales of order 10\,s to a few tens of
milliseconds) can plausibly be stirred up at all radii, but then
propagates towards the black hole as this is an accretion flow
\citep{lyu97,kot01,are06}. Propagation is governed by the local
viscous timescale, which strongly damps any faster variability. The
local viscous time is a function of radius, so the inner regions of
the flow can generate faster fluctuations than the outer radii,
coupling timescale and radii together. If the hot flow is also {\it
spectrally} stratified with radius, then this couples timescale and
spectrum together as well. This can quantitatively explain the
otherwise very puzzling observation that fluctuations in the
light curves of a hard and soft band are highly correlated, but with
the hard lagging behing the soft on a timescale which varies with the
variability timescale \citep{kot01}.

Several studies have looked at the energy spectrum of specific
features in the power spectrum (such as quasi-periodic oscillations,
QPOs) in BHB sources, but analyses of the broad-band variability are
much less frequent. The technique was pioneered by \citet{rev99}, who
looked at the energy spectra of Cyg X-1 in three different frequency
ranges. However, the focus of their study was to the iron line
features, whereas we use this to isolate the spectral shapes of the
broad band continuum components associated with different frequencies
of variability in the low/hard state of Cyg X-1.

We summarize the data and analysis technique in Sect.~\ref{analysis}
and present our results in Sect.~\ref{results}. Finally, we discuss
our findings in Sect.~\ref{discussion}.

\section{Data analysis}
\label{analysis}

Cygnus~X-1 is perhaps the most studied of all black hole binaries, with over 1000 observations
using the {\it RXTE} satellite alone. Its variability has been extensively reported in a number of
studies, spanning timescales from milliseconds to years \citep[e.g.,][]{rev00,rei02,zdz02}. Here we focus on
the broad-band variability in the 0.05--30\,Hz range seen in the hard spectral state. The fractional 
rms in this state is $\sim30\%$, and the power spectrum is well described by a combination of two
or more Lorentzian components. The peak frequency of these components vary on both long and
short timescales, and are correlated with spectral changes \citep[e.g.,][]{now00, pot03, axe05}. 

The study is performed using archival data of Cygnus~X-1 from the Proportional Counter Array \citep[PCA;][]{jah96} instrument onboard the {\xte} satellite, covering the energy range 3--35\,keV. Due to an antenna failure early on in the mission, most observations were performed using data modes where spectral resolution is lower to achieve high temporal resolution. To maximise the power of the analysis, we searched for all observations where the data mode B\_16ms\_64M\_0\_249 is available (a total of 8 ObsIDs), giving a good balance between spectral and temporal resolution. This mode allows us to extract 39 channels in the energy range 3 to 30\,keV with a time resolution of 16 ms, giving an upper limit of $\sim30$\,Hz for the power spectra. The lower range was chosen to be 0.01\,Hz. In the spectral fit, we also include data from the HEXTE instrument for the total spectrum, covering the energy range from 40 keV to 200 keV. 

All observations in the selected data mode are taken in the low/hard state,
with only small differences in luminosity and spectral shape. We 
choose the longest observation as this has the best signal-to-noise, taken on 
1996-03-30 (MJD 50172; ObsID 10238-01-05-000, 13ks exposure).
These data have been used in many previous studies, including the frequency-resolved analysis of \citet{rev06}.

\section{Results}
\label{results}

\subsection{Power spectra}

\begin{figure}
\includegraphics[width=8cm,angle=0]{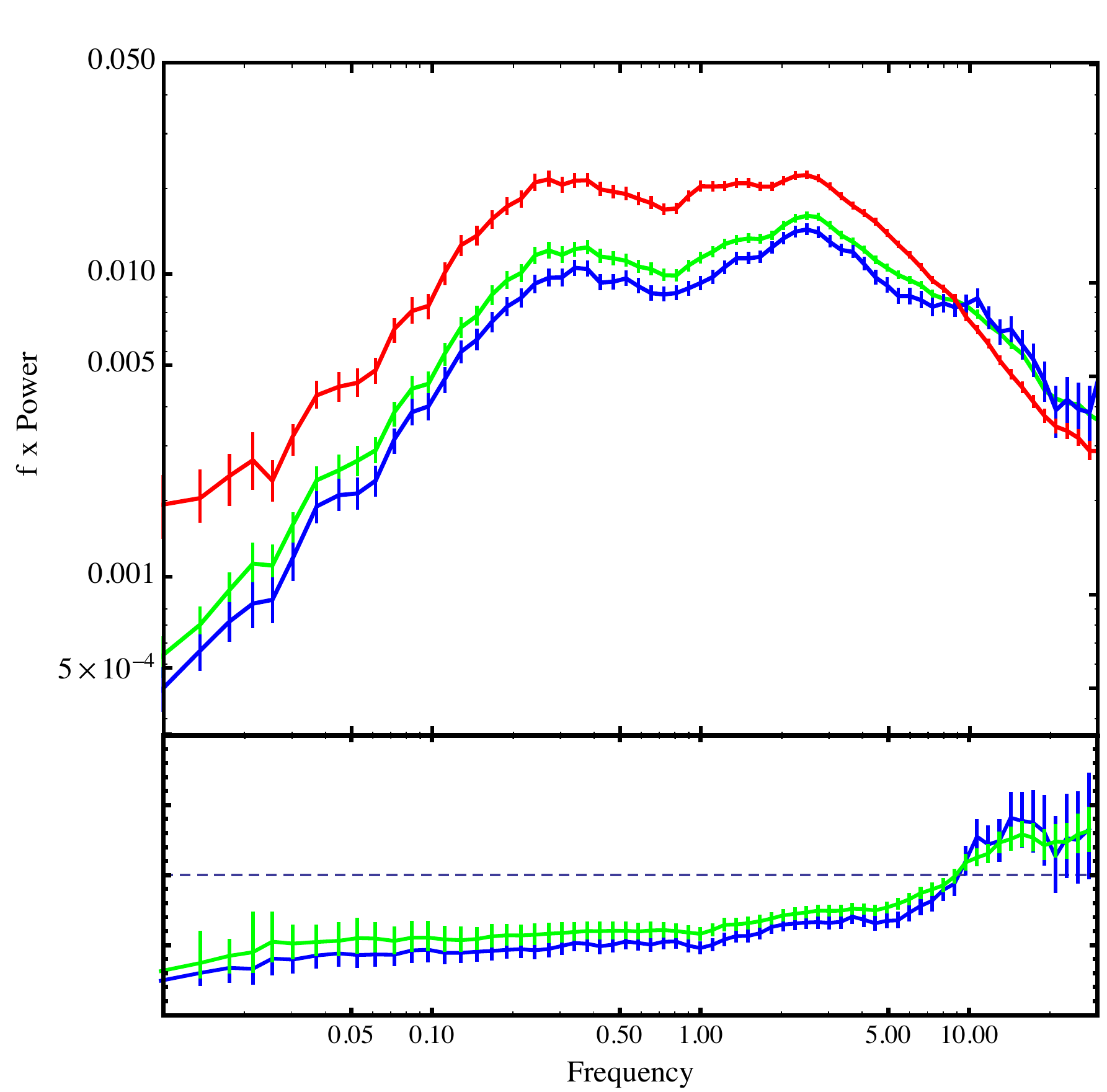}
\caption{{\it Upper panel:} Power spectra in three energy bands from ObsID 10238-01-05-000: 3-5 keV (red), 10-20 keV (green) and 20-35 keV (blue). {\it Lower panel:} Ratio of power of the 10-20 keV band (green) and 20-35 keV band (blue) compared to the 3-5 keV band.}
\label{pds}
\end{figure}

In the first step of the analysis, we extract the power spectrum in three broad energy ranges to get 
a view of its overall evolution with energy. The bands used for this were 3-5 keV (red), 10-20 keV(green)  
and 20-35 keV (blue) in Fig.~\ref{pds}. There
are two distinct features in the power spectrum, with excess of power around 
0.2~Hz, and 2~Hz. The overall shape of the power spectra are similar betwen the bands, except that the lowest energies have higher normalisation (and hence higher total rms power) except at the very highest frequencies. This can be seen more clearly  in the lower panel of Fig.~\ref{pds}, which shows the ratio of the power spectra in the mid and high energy bands to that at low energies.
The ratio switches from being fairly constant at low frequencies (less than 3~Hz), but then rising sharply to a much higher ratio above 10~Hz. This shows that there is a third feature in the power spectrum at high frequencies, but that this has a marked energy dependence, being much stronger at higher energies. 
Such a feature (and also a fourth component at even higher frequencies) has been reported in previous 
studies \citep[e.g.,][]{pot03}. 

We fit the three power spectra using three Lorentzian functions \citep[see][]{axe05}, and study the peak
frequency in each band. The results are presented in Table~\ref{freqtable}. We find that the frequencies do not show any significant energy dependence.

\begin{table}
\center
\begin{tabular}{l l l l}
 & $\nu_1$ (Hz) & $\nu_2$ (Hz) & $\nu_3$ (Hz) \\
 \hline
3--5\,keV & 0.29$\pm0.01$ & 2.3$\pm0.1$ & -- \\
10--20\,keV & 0.30$\pm0.01$ & 2.3$\pm0.1$ & 9.3$\pm1.3$ \\
20--35\,keV & 0.32$\pm0.02$ & 2.3$\pm0.1$ & 9.3$\pm0.9$ \\
\end{tabular}
\caption{Peak frequencies of the three Lorentzian components in the three broad energy bands.}
\label{freqtable}
\end{table}

\subsection{Time-averaged spectrum}

We first attempt to fit the time-averaged spectrum, using both PCA and HEXTE data, with a single Comptonisation region, 
similar to the fits in \citet{axe13}. This consists of 
thermal Comptonisation, parametrized using the model {\sc nthcomp} in {\sc Xspec}. We also include 
reflection ({\sc xilconv}) and relativistic smearing ({\sc kdblur}) of the Comptonised emission. For the smearing, the outer radius was frozen at 400 $r_g$ whereas the inner radius ($r_{\rm in}$) was left free. We characterise the interstellar absorption using 
use {\sc tbnew\_gas}, the updated version of {\sc tbabs} 
\citep{wil00}, with the assumption that all absorbing matter is neutral gas. The column density was frozen to 
$0.6\times10^{22}$\,cm$^{-2}$ \citep{now11}. We  also include a normalisation constant
between HEXTE and PCA, frozen at 0.9. Even if left free, the value does not change between models.
We do not include a thermal component from the disc as the data only cover energies above 3~keV, which is too high to include much emission from a typical low/hard state  temperature of 0.2~keV. We freeze the 
 input seed photons of our Comptonisation components to this
value \citep{ibr05}.

Our single-Comptonisation model is unable to provide a good fit to the data. This is not very 
surprising; as explained in Sect.~\ref{intro} many studies have pointed to inhomogeneous Comptonisation being present in {\cx1}. Following \citet{ibr05}, we therefore add a second thermal Comptonisation component, again with seed photon temperature frozen to 0.2~keV. 
In terms of {\sc xspec} components, the model becomes
{\sc tbnew\_gas$\times$(nthcomp+nthcomp+kdblur$\times$xilconv} {\sc $\times$(nthcomp+nthcomp))}.
Note that {\sc xilconv} is set to output only the reflected emission, and the parameters of its {\sc nthcomp} arguments
are tied to those of the direct emission.
This now gives an excellent fit to the data. Details of the two models are given in Table~\ref{fits1}.

\begin{table*}
\center
\begin{tabular}{l c c c c c c c c c c c}
Model & kT$_{\rm s}$ (keV) & $\Gamma_{\rm s}$ & norm$_{\rm s}$ & kT$_{\rm h}$ (keV) & $\Gamma_{\rm h}$ & norm$_{\rm h}$ & R & $\log\xi$ & r$_{\rm in}$ (r$_{\rm g}$) & $\chi^2$/dof \\
\hline
One Comp.  & -- & -- & -- & 480$^{+600}_{-160}$ & 1.68$^{+0.002}_{-0.002}$ & 1.86$^{+0.04}_{-0.04}$ & 0.37$^{+0.03}_{-0.03}$ & 3.30$^{+0.01}_{-0.01}$ & 45$^{+unc}_{-20}$ & 216.9/100\\
Two Comp.  & 1.8$^{+0.3}_{-0.3}$ & 2.01$^{+0.23}_{-0.10}$ & 0.56$^{+0.08}_{-0.21}$ & 116$^{+20}_{-14}$ & 1.66$^{+0.02}_{-0.01}$ & 1.94$^{+0.09}_{-0.10}$ & 0.27$^{+0.04}_{-0.04}$ & 2.98$^{+0.08}_{-0.12}$ & 7$^{+5}_{-2}$ & 52.2/97\\
\end{tabular}\\
\caption{Parameter values for the best fit of the total spectrum to the spectral model using one or two thermal
Comptonisation components. Subscripts s (soft) and h (hard) denote the two Comptonisation components, both of which have seed photons set to a 0.2~keV blackbody.} 
\label{fits1}
\end{table*}

We stress that while the two-component fit found here is the best fit using this model, it is not necessarily comparable to those found in other studies, and it is not unique. Indeed, previous analyses have found that a wide range of combinations of two Comptonisation regions are able to provide good fits to the spectrum of Cyg~X-1. Depending on the exact data set used, and using assumptions such as setting the same electron temperature for both regions, varying non-thermal electron fractions and different variations of the Comptonisation model, different conclusions are drawn about the properties of the Comptonisation \citep[see, e.g.,][]{ibr05,mak08,now11,mah18}. This divergent set of results underscores the degeneracy inherent in spectral analysis, stressing the need to consider also information from temporal analysis.

\subsection{Frequency resolved spectra}

\begin{figure}
\includegraphics[width=8cm,angle=0]{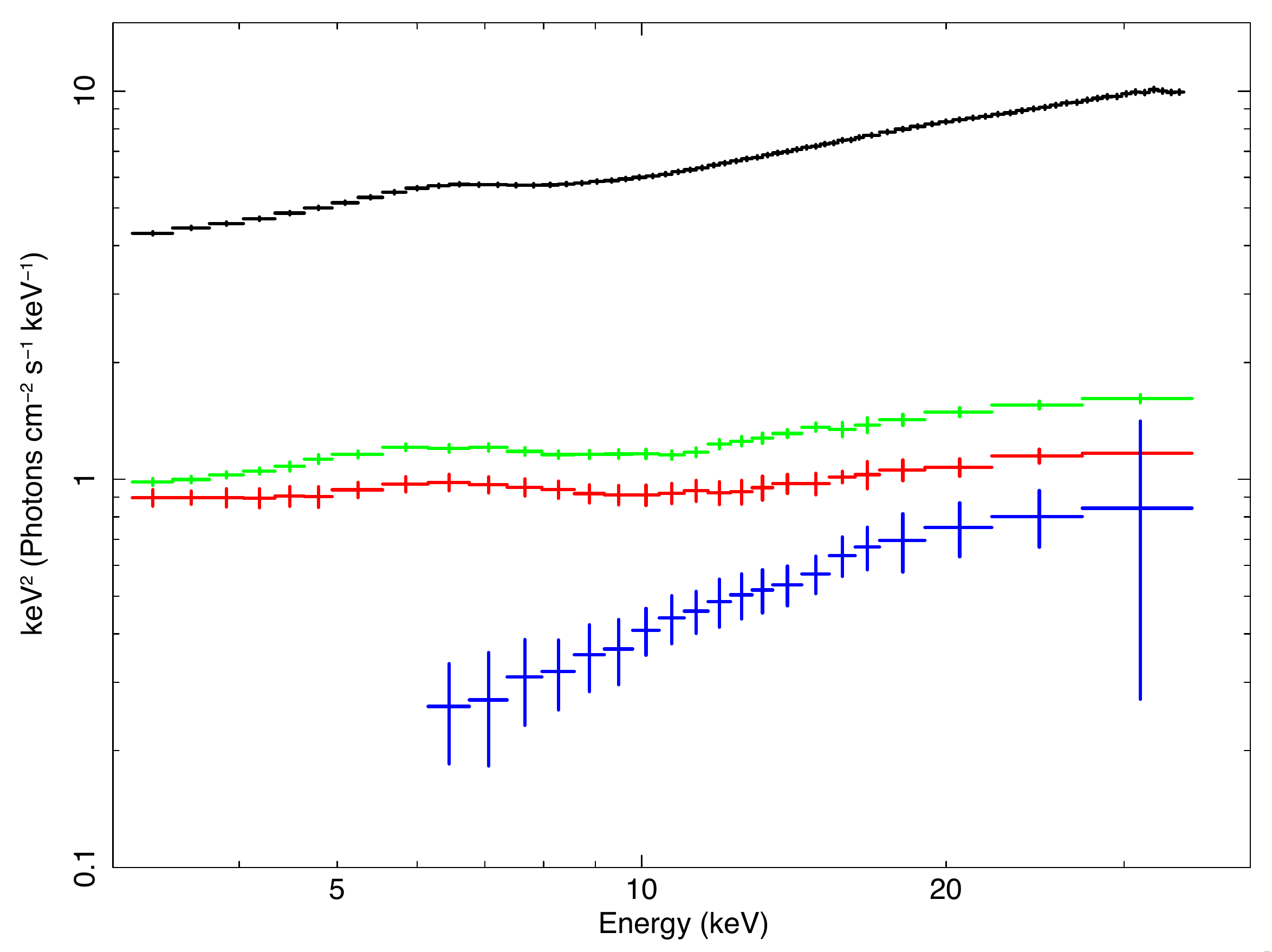}
\caption{{\it Upper panel:} Extracted spectra from the three components in the power spectrum. In order of rising
peak frequency: red ({\lone}), green ({\ltwo}) and blue ({\lthree}). The total spectrum (black) is also shown for comparison. While {\lone} and {\ltwo} show spectra similar to the total, {\lthree} is dramatically differerent.}
\label{05spec}
\end{figure}

In the next step we extracted the frequency-resolved spectrum using the same techniques
as in \cite{axe13}. We extract the light curve for every available energy channel and construct its power density spectrum. These were fit with 
three Lorentzian components (L$_1$, L$_2$ and L$_3$). We found that allowing the peak frequencies to vary in every channel led to 
very large uncertainties, and therefore froze the peak frequencies to the average of those found in Table~\ref{freqtable}: 
0.3, 2.3 and 9.3 Hz. This is justified since the frequencies did not change significantly between the energy bands (Table~\ref{freqtable}). With 
this approach we were able to constrain the components up to $\sim30$\, keV, above which the signal became too 
weak. In the first few channels, the highest frequency component is very weak, and its normalization is consistent 
with being zero. We integrate each Lorentzian in each energy channel to get an rms for that component at that energy, and multiply the time-averaged count rate by this value to extract frequency-resolved spectra corresponding to the noise components L$_1$ (red), L$_2$ (green) and L$_3$ (blue). These are presented 
along with the total spectrum in 
Fig.~\ref{05spec}, where all the data are shown deconvolved against a power law of index -2 (i.e., flat in the $\nu F_{\nu}$ representation) to aid in the comparison.

{\lone} and {\ltwo} (red and green points) show similar appearance to each other and to the total 
spectrum below 10~keV, including  features around the 6.4 keV iron line, indicating reflection. However, both are much softer than the total spectrum above 10 keV. This is consistent with the study
using {\it AstroSat} data presented by \citet{mis17}, who found that the lower frequency component weakens 
more at higher energies. We see the same trend in our data (Fig.~\ref{05spec}).

The most surprising behaviour is seen in the highest frequency component, {\lthree}. As expected, it is not 
detected at low energies, likely being too weak. However, at higher energies the spectrum is radically different 
from the other two, more resembling a single power law than the total spectrum. It is also significantly harder and has no obvious reflection features  \citep[see also][]{rev99}.

The data from the variability components were then individually fit, using the spectral model from the total
spectrum as template. The model was changed in steps until an acceptable fit was found. In the first step, the 
best-fit model for the total spectrum was merely scaled down, and tested against the data. Not surprisingly, this 
did not fit any of the variability components. {\lone} and {\ltwo} are softer than the total spectrum, whereas {\lthree} is 
too hard. In the second step, the relative normalisation between the two Comptonisation components was allowed to vary. This allowed us to find acceptable fits for {\lone} and {\ltwo}, but not for {\lthree}. 

As {\lthree} is much harder than the total spectrum, it is clear that the slope of the hard 
Comptonisation component must change to provide a good fit. We therefore allowed this parameter to vary.
However, we were still not able to find a good fit, unless we also let the seed photon temperature increase. 
Together, these two changes allowed us to find an acceptable fit to {\lthree} using a
single Comptonisation component; no reflection is required. It seems then that {\lthree}
provides us with a ``pure'' description of the hard Comptonisation!

\begin{figure}
\includegraphics[width=8cm,angle=0]{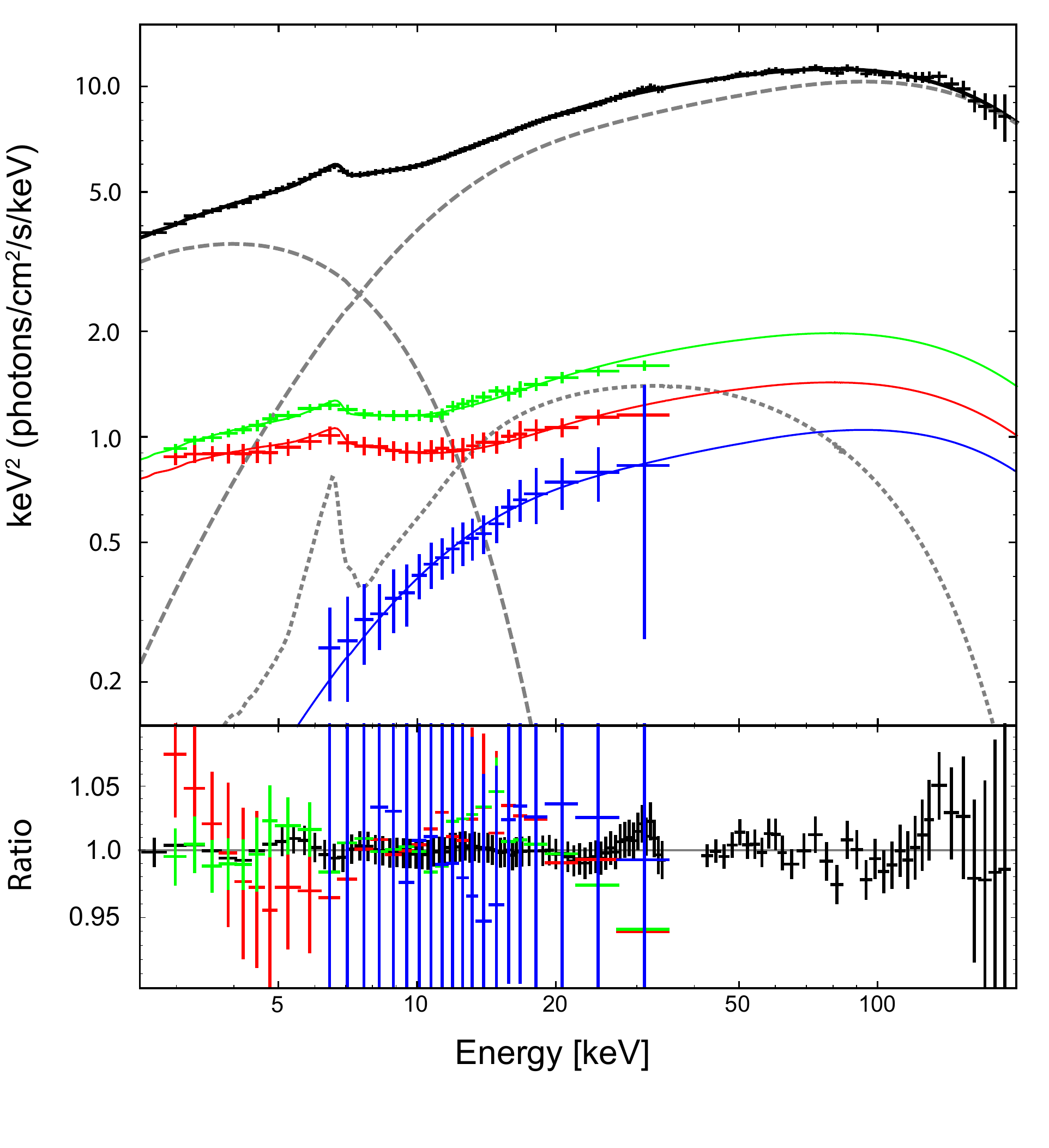}
\caption{Fits of the total spectrum (black points) and frequency-resolved spectra (red, green and blue 
points). The gray lines show the Comptonisation (dashed) and reflection (dotted) components in the total spectrum. Solid lines show the total spectrum for both the total spectrum as well as each individual frequency-resolved spectrum. The bottom panel shows the ratio between the data and fitted model.}
\label{allfit}
\end{figure}

In order to utilise this information, we now refit the total spectrum and {\lthree} together,
tying the hard Comptonisation in the total spectrum to the variability spectrum, and allowing for different electron
temperatures in the two Comptonisation components. The resulting fit quality is 
comparable to the one found before, but with the advantage that also {\lthree} can
be fit. As before, allowing the relative normalisation to vary also allows good fits to {\lone} and {\ltwo}. The results are 
presented in Table~\ref{partable2}, and the resulting spectra with residuals shown in Fig.~\ref{allfit}.

\begin{table*}
\begin{tabular}{l c c c c c c c c c c c}
Comp & kT$_{\rm e,s}$ & $\Gamma_{\rm s}$ & norm$_{\rm s}$ & kT$_{\rm seed,h}$ & kT$_{\rm e,h}$ & $\Gamma_{\rm h}$ & norm$_{\rm h}$ & R & $\log\xi$ & r$_{\rm in}$ & $\chi^2$/dof \\
 & (keV) &  &  & (keV) &  (keV) &  & $\times10^{-2}$ & & & (r$_{\rm g}$) & \\
\hline
Total  & 2.1$^{+0.2}_{-0.2}$ & 1.93$^{+0.03}_{-0.03}$ & 2.82$^{+0.11}_{-0.48}$ & 3.1$^\dagger$ & 145$^{+50}_{-32}$ & 1.69$^{+0.02}_{-0.03}$ & 1.8$^{+0.4}_{-0.6}$ & 0.24$^{+0.10}_{-0.05}$ & 3.00$^{+0.24}_{-0.19}$ & 5.1$^{+2.3}_{-1.5}$ & 61.6/171\\
{\lone}  & '' & ''  &  0.59$^{+0.02}_{-0.02}$ & '' & '' & '' & 0.22$^{+0.01}_{-0.01}$ & '' & '' & '' \\
{\ltwo} & '' & '' & 0.67$^{+0.01}_{-0.01}$ & '' & '' & '' & 0.31$^{+0.01}_{-0.01}$ & '' & '' & '' \\
{\lthree} & -- & -- & -- & 3.1$^{+0.5}_{-0.3}$ & '' & '' & 0.20$^{+0.03}_{-0.03}$ & -- & -- & '' \\
\hline
\end{tabular}\\
\raggedright
{\small $^\dagger$Value tied to that from {\lthree}.}
\caption{Parameter values for the best fit to the spectral model {\sc tbnew\_gas$\times$(nthcomp+nthcomp+kdblur
$\times$xilconv} {\sc $\times$(nthcomp+nthcomp))}. The total 
spectrum and the three variability components were all fit together. Subscripts s (soft) and h (hard) denote the two Comptonisation components.} 
\label{partable2}
\end{table*}

We note that the electron temperature of the soft Comptonisation component is almost comparable to
the seed photon temperature of the hard Comptonisation. Attempting to tie these parameters gives an acceptable
fit, yet slightly worse compared to those presented in Table~\ref{fits1}. We therefore leave them separated. 

The parameters in Table~\ref{partable2} also allow us to look at the ratio between normalisation parameters for the soft and hard Comptonisation components. This ratio (norm$_{\rm s}$/norm$_{\rm h}$) is 268 for {\lone} and 216 for {\ltwo}, making it clear that the hard Comptonisation is relatively stronger in {\ltwo} than in {\lone}. The same ratio is 112 in the total spectrum, indicating that the hard Comptonisation is less pronounced in the variability spectra compared to the total spectrum. 

\begin{figure}
\includegraphics[width=8.4cm,angle=0]{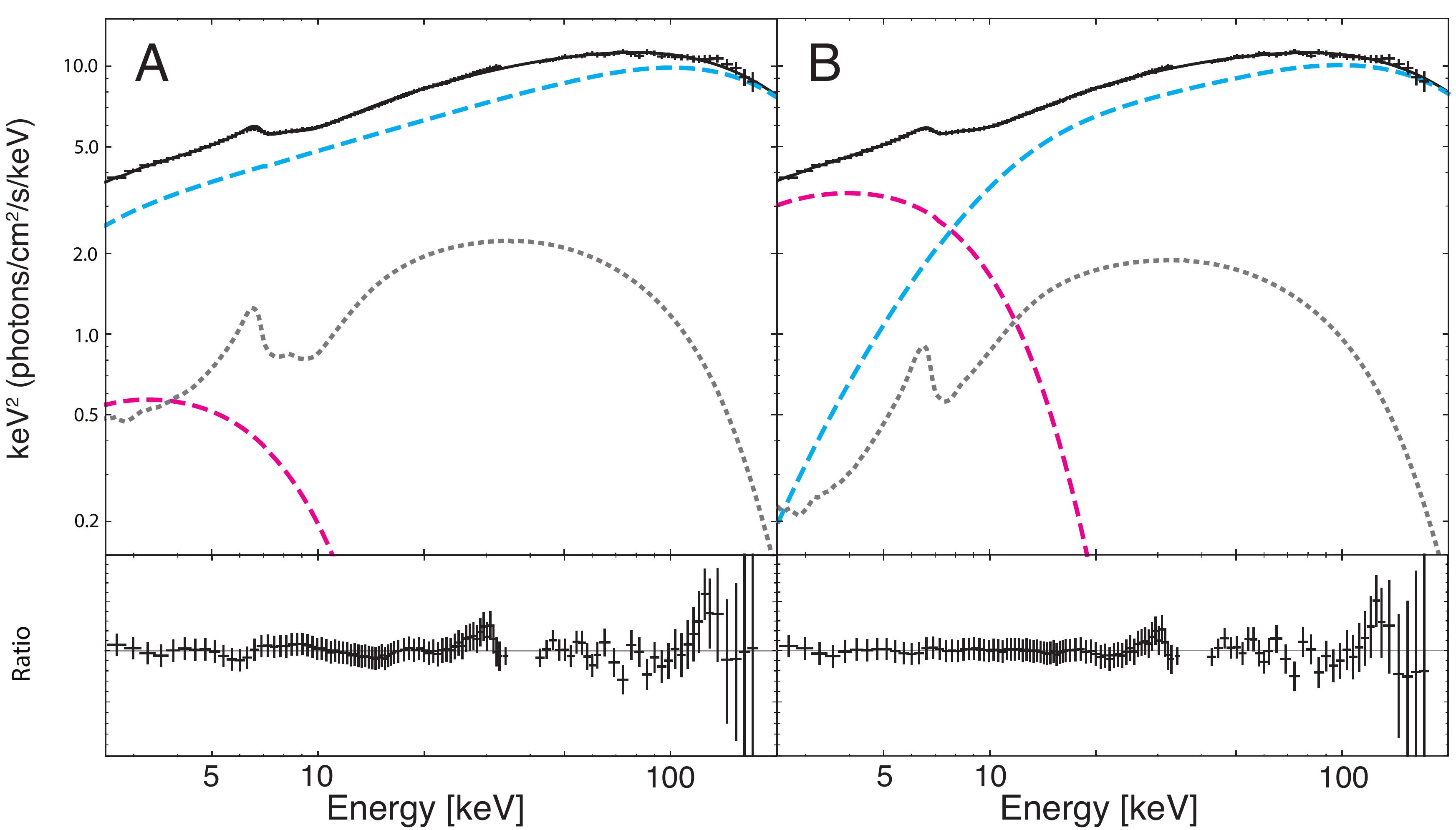}
\caption{Comparison of spectral fits using different seed photon temperature for the hard Comptonisation component. {\bf A}: Original model using kT$_{\rm seed,h}$ = 0.2\,keV. $\chi^2$/dof = 52.2/97. {\bf B}: Model using kT$_{\rm seed,h}$ = 3\,keV, as indicated by the variability spectra. $\chi^2$/dof = 34.7/97. Upper panels show the data and model components: soft Comptonisation (magenta), hard Comptonisation (cyan) and reflection (gray). Lower panels show the ratio between data and model.}
\label{twofits}
\end{figure}

Finally, we compare the best-fit model found using the variability components with that presented in Table~\ref{fits1}. The main difference between the models are that the seed photon temperature of the hard Comptonisation component is no longer set to 0.2 keV. Instead, we fix it at 3 keV, as indicated by the spectrum of {\lthree}. Surprisingly, the resulting fit proves to be a significant improvement over the result in Table~\ref{fits1}, with a $\chi^2$ of 34.7 compared to 52.2. Figure~\ref{twofits} shows a direct comparison between the two fits. Although both models give a good fit, using a higher seed photon temperature gives a significantly better fit statistic, and the PCA residuals improve. The model found by considering also the temporal information has thereby also led to a better global fit. This further strengthens the result that the two Comptonisation regions are spatially distinct, with the harder region seeing much more energetic seed photons.

The high seed photon temperature required by the data implies that the innermost region (where {\lthree} arises) is shielded from the disc photons. This in turn suggests that Comptonised emission arising in this region would not reach the disc, and thereby not be reprocessed. Unfortunately, the spectrum of {\lthree} does not allow us to test this prediction, as the uncertainties on the data are large  (cf. the direct emission and reflection in Fig~\ref{twofits}B). Allowing only the low-temperature Comptonisation component to be reflected in the total spectrum gives an acceptable but worse fit compared to the one in Table~\ref{partable2} ($\chi^2$/dof of 74.3/171 compared to 61.1/171). We therefore choose to keep the fit where both Comptonisation components are reflected. However, the worsening of the fit when considering the shielding effect of the softer component could be an indication that there are details in the data which our two-component model is unable to capture. We try to constrain these additional features in a model independent way below. 

\subsection{Isolating different radii}

\begin{figure}
\includegraphics[width=8cm,angle=0]{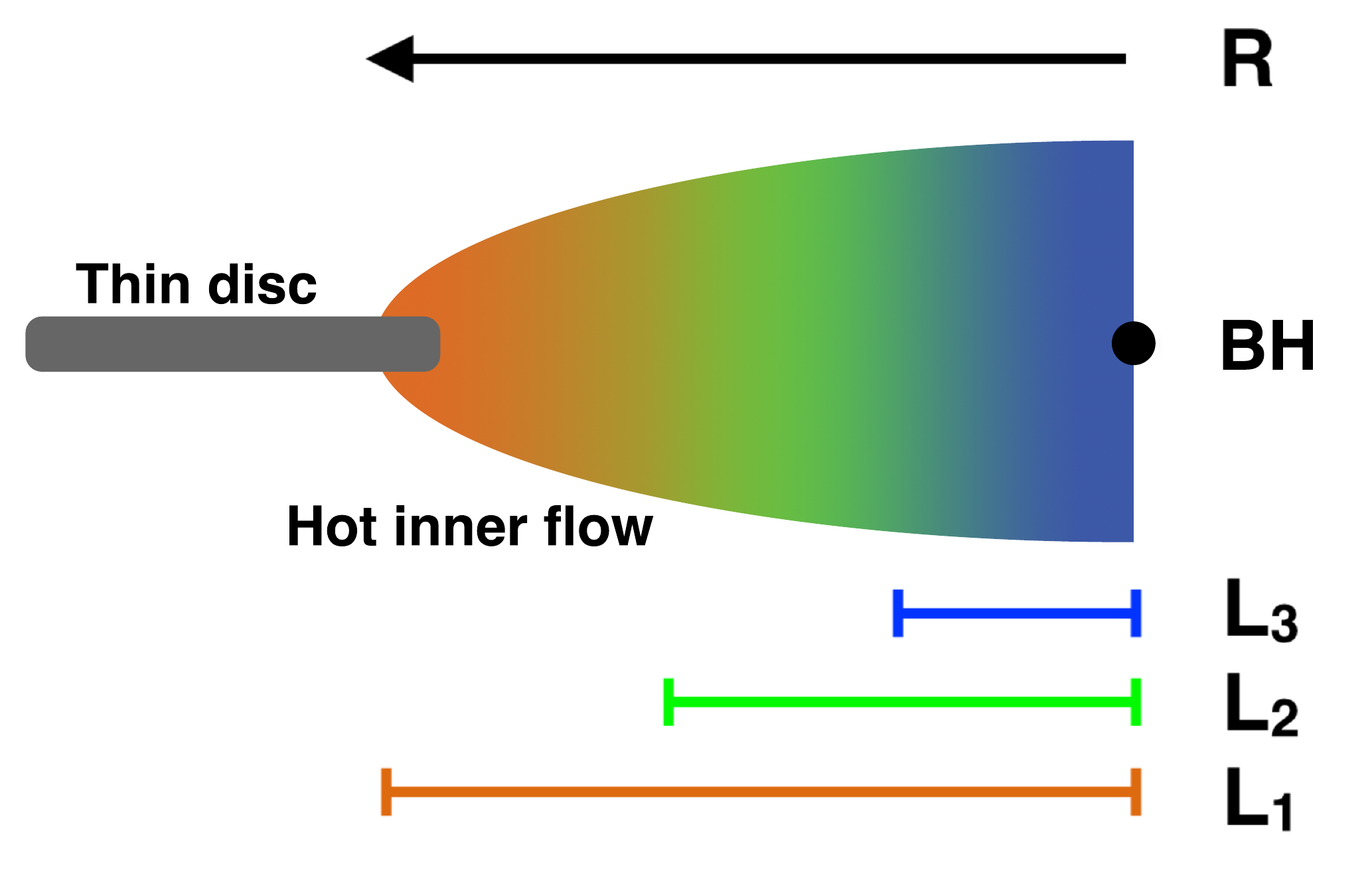}
\caption{Cartoon picture showing the disc (grey) and regions of the inner accretion flow contributing to each variability component.}
\label{flucmodel}
\end{figure}

As described in Sect.~\ref{intro}, the broad-band variability likely arises due to propagating fluctuations in the
accretion flow. In this picture, slower propagations arise in the outer regions of the flow, and as they propagate
inwards, more rapid fluctuations are added. Variability at a given frequency is produced at a characteristic radius, so includes the spectrum produced at that radius, but then it propagates inwards so it is the sum of different spectral components at all radii interior to that where the characteristic frequency of the component is generated. This is shown schematically shown in Fig.~\ref{flucmodel}, where the sketch
is of a gradually stratified flow, with each region contributing a harder spectrum as the radius decreases, rather than the two component model used above. 

The cartoon
image of Fig.~\ref{flucmodel} reveals a model-independent way to extract the spectra as a function of radius. By studying
the {\it difference} between the variability spectra, we can attempt to isolate the spectrum which is characteristic of the radius at which each variability component is produced. 

\begin{figure}
\includegraphics[width=8cm,angle=0]{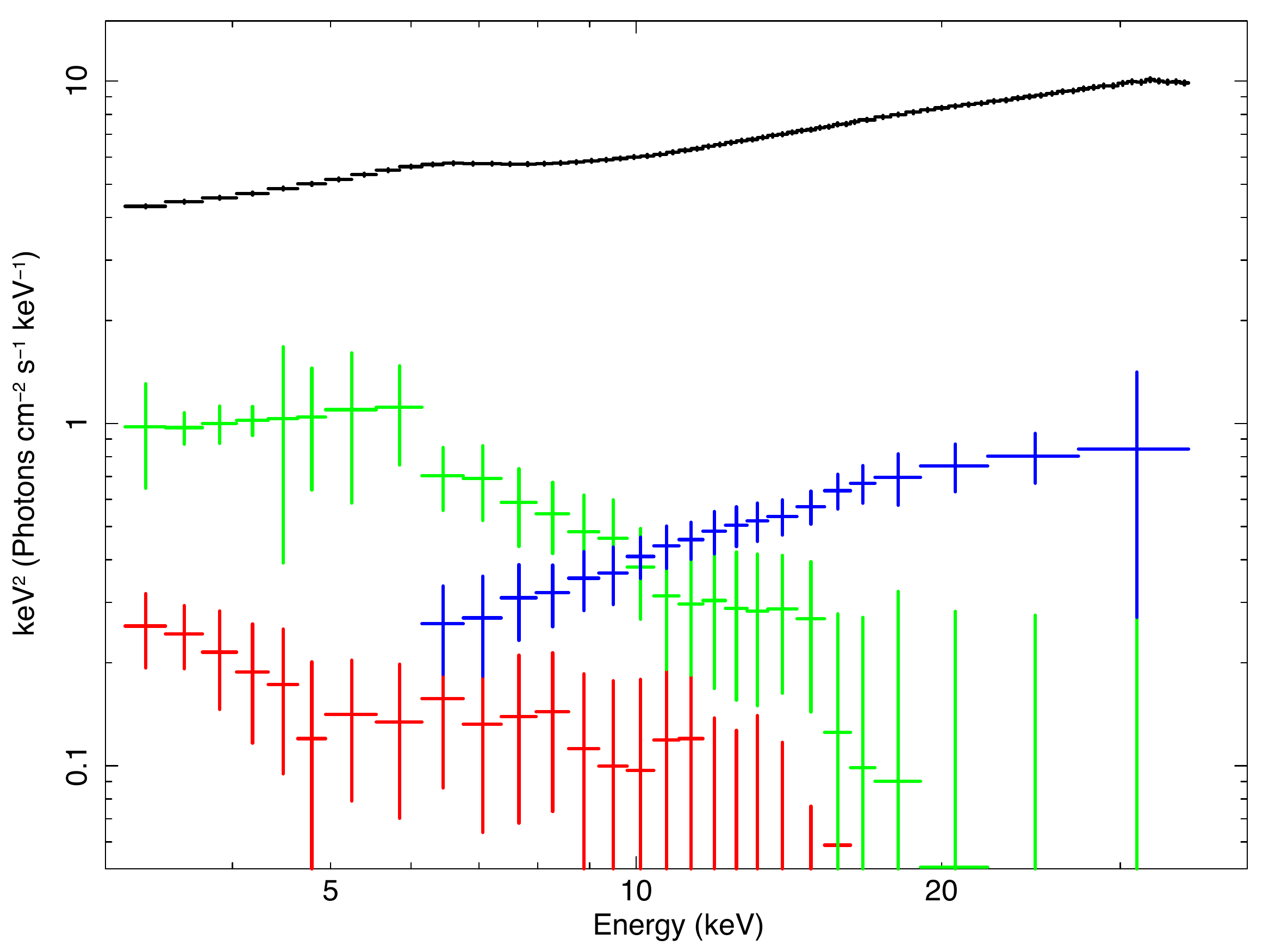}
\caption{Spectra of each variability component after subtracting the contribution from higher frequencies (smaller
radii). The colors are the same as in Fig.~\ref{05spec}}
\label{subspec}
\end{figure}

As described above, the highest frequencies are expected to be the ``cleanest''. This is supported by our result that the spectrum of {\lthree} is well-fit by a single pure Comptonisation component. {\ltwo} has a lower frequency, and it should therefore contain the same contributions as {\lthree}, as well as emission from larger radii. To isolate the emission in the regions successively further out, we rescaled the {\lthree} spectrum to match that from {\ltwo} in the highest energy bin (where the contribution is dominated by the innermost region), and then subtracted the {\lthree} spectrum. This is the maximum contribution from the hard innermost region, so this spectra is the softest that can be characteristic of the region where {\ltwo} is produced. 
Similarly, we rescaled and subtracted the spectrum of {\ltwo} from {\lone}. The results are shown in Fig.~\ref{subspec}.

While the data points have large uncertainties, some general features can still be distinguished. 
The spectrum characteristic of the largest radii (red) is the softest, and is subtly different to the shape of the soft Comptonisation in our two-component fits. The spectrum which can be associated with {\ltwo} (green) is appreciably harder than that of {\lone}, while that of {\lthree} is well described by a single hard Comptonisation component with high electron temperature and seed photons around a few keV (as above). 

These results suggest a more stratified region than in our two component model fits. Unfortunately, the data do not allow us to explore this further. However, it is clear that  the technique has the potential to reveal even greater details of the stratified flow than shown in this study when combined with new data from {\it NICER} and/or {\it AstroSat}. 

\section{Discussion}
\label{discussion}

It has been known for a long time that there are more than two components in the broad-band variability
spectrum of Cyg~X-1 \citep[e.g.,][]{now00,pot03}. However, the frequency range studied here (0.01-30 Hz)
is dominated by two components, with a third entering only in the hardest observations \citep{axe08}. It is
therefore interesting to see the presence of this third component in the higher energy bands, even when
it is not evident in the lowest band. 

Previous results have shown that the variability components change in response to spectral evolution of the 
source, becoming weaker and moving to higher frequencies as the hardness decreases 
\citep{pot03,axe05,axe06,gri14}. Our results indicate that there is also an energy dependence present.
Because the frequencies do not change between the energy bands, it is likely that the third variability
component appears due to increased relative strength as the energy increases - this is additionally supported 
by its hard spectrum. Having the
strength of the component vary with energy also explains the result of \citet{now00}, who found that the 
best-fit combination of Lorentzian functions matching the power spectrum was dependent on energy.

In the spectral fits, we again confirm previous results that a single Comptonisation region is not sufficient
to fit the spectrum of Cyg~X-1. This is a general feature in many BHB systems \citep[e.g.,][]{ibr05,mak08,yam13,hja16}. This result is further strengthened by the addition of the frequency-resolved data. While it is clear that both {\lone} and {\ltwo} show signs of Comptonisation, they are both softer than the total spectrum. There is thus no way that a single Comptonisation spectrum can explain both their spectra and the total. The point is driven home by {\lthree}, which appears to match a clean Comptonisation spectrum, yet this spectrum is much too hard to explain the softer spectra. The only alternative is to allow multiple Comptonisation regions.

If there are indeed two (or more) separate Comptonisation components, as suggested by both the spectrum and the 
variability, they cannot be completely independent. If the two components would vary randomly with respect
to each other, the rms is expected to be lower in the energy range where they contribute equally, as opposed
to where one component dominates. There is no indication of such behaviour, either in the observations studied
here or in previous works \citep[e.g.,][]{rev01}. 

The smooth evolution of the rms instead supports the view of variability propagating through the flow, with each
radius contributing most strongly at frequencies close to the local viscous timescale. Slower variations from the
outer regions then modulate the faster variability, naturally coupling the variability at different timescales. This 
picture also couples higher frequencies to regions closer to the black hole, predicting that the energy spectrum 
of the variability components will get harder with increasing frequency. This matches the observations reported
here, as well as many previous results. 
 
While the slight differences between the first two variability components can be explained through sampling of
different radii in a smoothly stratified flow, the third component is radically different, with seed photons as well as 
electrons being very energetic. 
As it is tied to the highest frequencies, it likely originates in the innermost part of the flow. The spectral shape then points to this region being shielded from the soft seed photons from the disc by the
softer Comptonisation regions.  Hence it should also have no reflection, as observed, although the uncertainties are large. 

A further clue to the radial spectral stratification comes from subtracting the different frequency resolved spectra.  Assuming the propagation framework is generally correct, these should mirror spectral components representative of the region where the variability component arises. While we are not able to directly fit any specific emission component to these spectra, the data by themselves indicate that {\ltwo} arises in a higher-energy environment than {\lone}. Both components resemble Comptonised emission rather than a blackbody, pointing to an optically thin flow rather than the accretion disc itself, consistent with a truncated disc geometry.

We can now put all our results in the context of the accretion geometry. The similarity of the subtracted spectra to Comptonisation leads us to place the origin of the variability components in the hot inner flow. {\lone} has both lowest frequency and the softest spectrum, and is thereby likely to arise furthest out in the flow. The spectrum of {\ltwo} is slightly harder, placing it at smaller radii. However, both these variability spectra show clear signs of reflection (such as a noticeable iron line), indicating that they cannot arise too far away from the thin, optically thick disc. In contrast, {\lthree} has a very hard spectrum which shows no sign of reflection, placing its origin well inside the disc truncation radius and close to the innermost regions of the flow. The seed photon temperature of several keV also makes it likely that they come from Comptonisation in the outer parts of the hot flow, rather than directly from the disc. Combining emission from all these regions gives the total spectrum.

Even if we can paint a generally consistent picture, there are still questions which remain unanswered. For instance, a mechanism must be found to explain the frequencies of the variability components. If variability is created at all frequencies, why are these ones picked out? The most obvious answer is that the variability components are tied to certain radii in the flow, yet this does not match the fundamental assumption in propagation model, where fluctuations generated at one radii will propagate and thereby be present at all smaller radii. To prevent this, a mechanism must be invoked to allow one region to dominate, for example by assuming increased emissivity at certain radii (perhaps as a response to increased turbulence) or causing fluctuations to dampen as they propagate. However, our current results do not allow us to test any such hypotheses.

While the results reported here constitute an important step in breaking the spectral degeneracy, it is clear that there are many issues left to solve, both in the propagating fluctuations framework and to understand these data. For example, while \citet{rap17b} were able to find good agreement between model and observations for power spectra, time lags and coherence as a function of energy in Cyg~X-1 using propagating fluctuations, the same analysis showed discrepancies when applied to the BHB XTE~J1550-564 \citep{rap17a}. It is thus not surprising that also our results require the model to be refined in order to explain all observed features; indeed, such efforts are already underway (Mahmoud et al., in prep.). Equally important are more observational results, allowing us to look for systematic changes, for example as a function of spectral state. Such analysis is in progress, and will be reported in a future paper (Axelsson et al., in prep.).

\section{Summary and Conclusions}

We have analysed the broad-band variability of {\cx1} in the low/hard state using frequency-resolved spectroscopy. We find that the two main power spectral components, {\lone} and {\ltwo}, 
have spectra which 
are too soft to match the total spectrum at higher energies. A third variability 
component appears at higher frequencies 
in the power spectrum, and this component has a drastically harder 
spectrum. This is mainly due to its seed photon energy being substantially higher than those seen by the flow at larger radii. These results can be interpreted in terms of propagating fluctuations
through a Comptonisation clould which is radially stratified. The outer regions see the seed photons from the disc, and Compton cool on them, producing soft spectra. These regions also shield the inner parts of the flow from direct illumination by the disc as they are (moderately) optically thick. We model this directly using only two Comptonisation components, but subtraction of the spectral components indicates that there may be more gradual radial stratification of the outer region. This shows the potential of better data to directly deconvolve the radial properties of the hot inner flow, and hence determine whether this contains the imprint of the radius from which the compact jet is launched and powered. 

\section*{Acknowledgments}
This work was supported by The Carl Trygger Foundation (grant CTS 16:41) and Ivar Bendixsons foundation, and has made use of data obtained through the High Energy 
Astrophysics Science Archive Research Center (HEASARC) Online Service, provided by NASA/Goddard Space Flight Center. CD acknowledges STFC funding under grant ST/L00075X/1 and a JSPS long term fellowship L16581.

\end{document}